\documentclass[11pt,showpacs,preprintnumbers,amsmath,amssymb,prd,nofootinbib,superscriptaddress]{revtex4-2}

\usepackage{dcolumn}
\usepackage{bm}
\usepackage{ifpdf}
\usepackage{hyperref}
\usepackage{bm}
\usepackage{xcolor,color,graphicx,graphics}
\usepackage[spanish,english]{babel}
\usepackage[latin1]{inputenc}
\usepackage[OT1]{fontenc}
\usepackage{latexsym,amssymb,amsmath,amsfonts}
\usepackage{makeidx}
\usepackage{epsfig,subfigure}
\usepackage{natbib}
\usepackage{epstopdf}
\usepackage{mathrsfs}
\usepackage{hyperref}
\hypersetup{colorlinks=true, linkcolor=blue, citecolor=blue, urlcolor=blue}
\usepackage{enumerate}

\everymath{\displaystyle}
\usepackage{graphicx}

\usepackage[T1]{fontenc}
\usepackage{amsmath}
\usepackage{amssymb}
\usepackage{graphicx}
\usepackage{xcolor}

\newcommand{\bea}{\begin{eqnarray}}
\newcommand{\eea}{\end{eqnarray}}

\newcommand{\orcid}[1]{\href{https://orcid.org/#1}{\includegraphics[width=10pt]{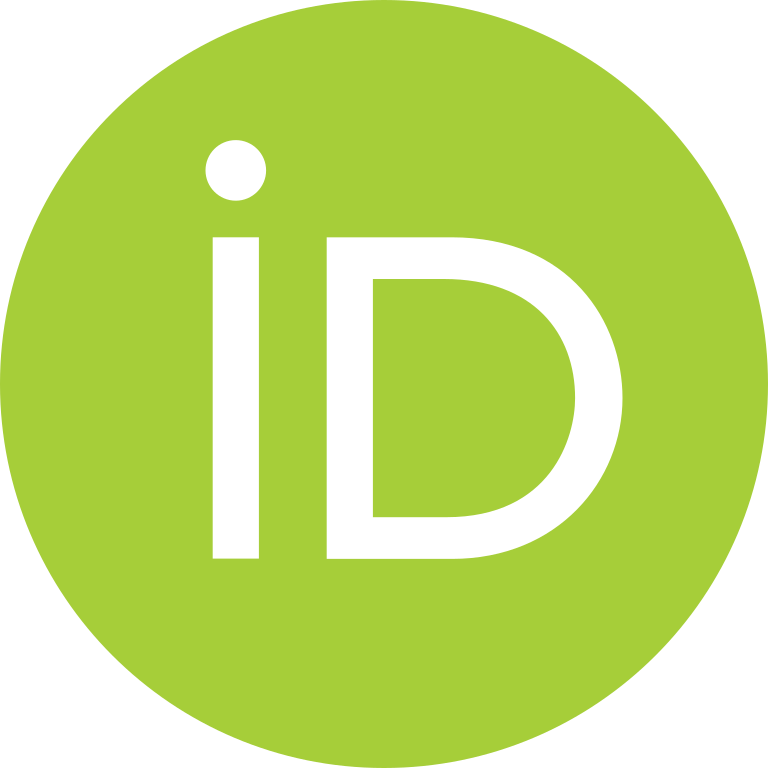}}}

\usepackage{fixmath}

\begin{document}

\title{Thermal Casimir effect in G\"{o}del-type universes}

\author{A. F. Santos \orcid{0000-0002-2505-5273}}
\email{alesandroferreira@fisica.ufmt.br}
\affiliation{Instituto de F\'{\i}sica, Universidade Federal de Mato Grosso,\\
78060-900, Cuiab\'{a}, Mato Grosso, Brazil}

\author{Faqir C. Khanna \orcid{0000-0003-3917-7578} \footnote{Professor Emeritus - Physics Department, Theoretical Physics Institute, University of Alberta\\
Edmonton, Alberta, Canada}}
\email{fkhanna@ualberta.ca; khannaf@uvic.ca}
\affiliation{Department of Physics and Astronomy, University of Victoria,\\
3800 Finnerty Road, Victoria BC V8P 5C2, Canada}

\begin{abstract}

In this paper, a massless scalar field coupled to gravity is considered. Then the Casimir effect at finite temperature is calculated. Such development is carried out in the Thermo Field Dynamics formalism. This approach presents a topological structure that allows for investigating the effects of temperature and the size effect in a similar way. These effects are calculated considering G\"{o}del-type solutions as a gravitational background. The Stefan-Boltzmann law and its consistency are analyzed for both causal and non-causal G\"{o}del-type regions. In this space-time and for any region, the Casimir effect at zero temperature is always attractive. However, at finite temperature, a repulsive Casimir effect can emerge from a critical temperature.

\end{abstract}

\maketitle

\section{Introduction}

The Casimir effect is a quantum remarkable phenomenon with numerous applications. This effect was first proposed by H. Casimir, in 1948 \cite{Casimir}. It describes an attractive force that arises between two parallel conducting plates placed in the vacuum of a quantum field. About ten years after the theoretical proposal, experimental confirmation was carried out \cite{Sparnaay}. Recently, the experimental accuracy has increased significantly \cite{Lamoreaux, Mohideen, Bressi, Most, Kim}. The original idea was developed using the electromagnetic field. However, nowadays this phenomenon appears in any quantum field. The Casimir effect emerges when boundary conditions or topological effects are imposed on a quantum field. As a consequence, the vacuum energy of the field is modified \cite{Milton, Milonni, Bordag}. In this paper, the quantum field to be considered is the massless scalar field coupled to gravity. In this context, thermal effects are investigated having as a gravitational background the G\"{o}del-type solutions.

Temperature changes the properties and behavior of any system. Furthermore, phenomena at zero temperature generally do not occur in nature. To be more realistic let us introduce the temperature effect in our theory. There are different approaches in the literature to introduce temperature effects into a quantum field theory \cite{Matsubara, Schwinger}. Here the Thermo Field Dynamics (TFD) formalism  \cite{Umezawa1, Umezawa2, Kbook, Umezawa22, Khanna1, Khanna2} is considered. TFD is a thermal quantum field theory that exhibits a topological structure. This topology is defined as $\Gamma_D^d=(\mathbb{S}^1)^d\times \mathbb{R}^{D-d}$, where  $D$ are the space-time dimensions and $d$ is the number of compactified dimensions. This implies that any set of dimensions of the manifold can be compactified in a circumference  $\mathbb{S}^1$.  Due to this characteristic, different effects such as the Stefan-Boltzmann law and Casimir effect at zero and non-zero temperatures can be calculated in this formalism in the same way, just considering different compactifications along the space-time dimensions. In this work, the TFD formalism is used to calculate the Stefan-Boltzmann law and Casimir effect on a gravitational background described by the G\"{o}del-type universe.

In 1949, Kurt G\"{o}del proposed a cosmological model that is a solution to Einstein's equations \cite{Godel}. It is a rotating cosmological model with a non-vanishing cosmological constant and a dust-like matter source. The main feature of this solution is the possibility of the existence of Closed Time-like Curves (CTCs) that lead to the breakdown of causality. Violation of causality is not a unique characteristic of the G\"{o}del solution, there are other solutions from the general theory of relativity that allow such CTCs \cite{Stockum, Gott}. This metric has been generalized to the so-called G\"{o}del-type metric \cite{Reboucas}. Various properties of this metric have been investigated \cite{Reboucas1, Reboucas2, Reboucas3}. In the G\"{o}del-type solution there is a specific relationship between two parameters that allow the investigation of three classes of solutions that lead to three different regions: (i) non-causal; (ii) causal and (iii) there is an infinite sequence of alternating causal and non-causal regions. Here, the G\"{o}del-type universe is considered to calculate the Stefan-Boltzmann law and Casimir effect at finite temperature. The Casimir effect in the G\"{o}del space-time has been investigated \cite{Kho, our}. However, these works do not analyze what happens in the energy density and in the Casimir effect in the causal and non-causal G\"{o}del-type regions. Therefore, here it is investigated whether there is an influence of these regions on these phenomena.

This paper is organized as follows. In section II, the theory is presented. The vacuum expectation value of the energy-momentum tensor associated with the massless scalar field coupled to gravity is calculated. In section III, a brief introduction to the TFD formalism is given. In order to obtain physical quantities, the energy-momentum tensor is rewritten, where a renormalization procedure is performed. In section IV, some characteristics of the G\"{o}del-type universe are explored. In section V, thermal applications on a G\"{o}del-type background are investigated. Different topologies are chosen, then thermal effects and the size effects are calculated in this cosmological model. In section VI, some concluding remarks are made.

\section{The theory: Scalar field coupled to gravity}

Here the theoretical model describes the gravitational field coupled with a massless scalar field. Its Lagrangian is given as \cite{Birrel}
\bea
{\cal L}=\frac{1}{2}\left(g^{\mu\nu}\partial_\mu\phi(x)\partial_\nu\phi(x)-\xi R \phi(x)^2\right),\label{1}
\eea
where $g^{\mu\nu}$ is the metric tensor, $\phi(x)$ is the massless scalar field, $R$ is the Ricci scalar and $\xi$ is the coupling constant. The main objective of this work is to investigate this model in the G\"{o}del-type universe at finite temperature. To develop such a study, the energy-momentum tensor associated with Eq. (\ref{1}) must be calculated. Using the definition, 
\bea
T_{\mu\nu}=-\frac{2}{\sqrt{-g}}\frac{\delta {\cal L}}{\delta g^{\mu\nu}},
\eea
the energy-momentum tensor is given as
\bea
T_{\mu\nu}(x)&=&\frac{1}{2}g_{\mu\nu}\partial^\rho\phi(x)\partial_\rho\phi(x)-\partial_\mu\phi(x)\partial_\nu\phi(x)+\xi\left(R_{\mu\nu}-\frac{1}{2}g_{\mu\nu}R+g_{\mu\nu}\Box-\partial_\mu\partial_\nu\right)\phi(x)^2,
\eea
with $R_{\mu\nu}$ being the Ricci tensor and $\Box=g^{\mu\nu}\partial_\mu\partial_\nu$ is the d'Alembertian operator. 

It is important to observe that, due to the product of two fields at the same space-time point, the energy-momentum tensor becomes a divergent quantity. In order to obtain a finite quantity, this tensor is written at different points in space-time. This is a well-known technique used in quantum field theory, for examples see  \cite{Chris, Chris1, Deca}.  Then
\bea
T_{\mu\nu}(x)&=&\lim_{x'\rightarrow x}\Bigl\{\Delta_{\mu\nu}\tau\left[\phi(x)\phi(x')\right]-\Sigma_{\mu\nu}\delta(x-x')\Bigl\},
\eea
where $\tau$ is the time ordering operator and
\bea
\Delta_{\mu\nu}&=&\frac{1}{2}g_{\mu\nu}\partial^\rho\partial'_\rho-\partial_\mu\partial'_\nu+\xi\left(R_{\mu\nu}-\frac{1}{2}g_{\mu\nu}R+g_{\mu\nu}\Box-\partial_\mu\partial'_\nu\right)\\
\Sigma_{\mu\nu}&=&-\frac{i}{2}g_{\mu\nu}n_0^\rho\,n_{0\rho}+in_{0\mu}n_{0\nu}.
\eea
Here $n_0^\mu=(1,0,0,0)$ is a time-like vector and the canonical quantization for the scalar field, i.e. $[\phi(x),\partial'^\mu\phi(x')]=in_0^\mu\delta({\vec{x}-\vec{x'}})$, has been used. 

To make the proposed applications, the vacuum expectation value of the energy-momentum tensor is calculated, i.e.
\bea
\left\langle T_{\mu\nu}(x)\right\rangle&=&\lim_{x'\rightarrow x}\Bigl\{i\Delta_{\mu\nu}G_0(x-x')-\Sigma_{\mu\nu}\delta(x-x')\Bigl\},\label{VEV}
\eea
where the definition of the  massless scalar field propagator
\bea
\left\langle 0\left|\tau[\phi(x)\phi(x')]\right| 0 \right\rangle=iG_0(x-x')
\eea
has been considered.

In order to use Eq. (\ref{VEV}) for some applications at finite temperature, an introduction to TFD formalism is presented in the next section.

\section{TFD formalism}

Thermo field dynamics formalism is a  real-time approach that introduces the effects of temperature into a quantum field theory without losing the temporal evolution of the system. This formalism is built on ideas where the statistical average of an arbitrary operator is equal to the vacuum expectation value in a thermal vacuum. To construct a thermal vacuum, two elements are needed: (i) the doubling of the Hilbert space and (ii) the Bogoliubov transformation. The duplicated Hilbert space ${\cal S}_T={\cal S}\otimes \tilde{\cal S}$, also known as the thermal Hilbert space, consist of the original Hilbert space ${\cal S}$ and the dual (tilde) Hilbert space $\tilde{\cal S}$. The map between the spaces tilde and non-tilde is defined by the tilde (or dual) conjugation rules, which are defined as
\bea
(A_iA_j)^\thicksim & =& \tilde{A_i}\tilde{A_j}, \quad (\tilde{A_i})^\thicksim = -\eta A_i,\\
(A_i^\dagger)^\thicksim &=& \tilde{A_i}^\dagger, \quad\quad (cA_i+A_j)^\thicksim = c^*\tilde{A_i}+\tilde{A_j},\nonumber
\eea
with $\eta = -1$ for bosons and $\eta = +1$ for fermions. The other basic element of the TFD formalism is the Bogoliubov transformation, which introduces a rotation in the tilde and non-tilde operators. In this way, the temperature effect emerges from a condensed state. For an arbitrary operator ${\cal A}(k)$, this transformation is given as
\bea
\left( \begin{array}{cc} {\cal A}(k, \alpha)  \\\eta\, \tilde {\cal A}^\dagger(k,\alpha) \end{array} \right)={\cal B}(\alpha)\left( \begin{array}{cc} {\cal A}(k)  \\ \eta\,\tilde {\cal A}^\dagger(k) \end{array} \right),
\eea
where  ${\cal B}(\alpha)$ is defined as
\bea
{\cal B}(\alpha)=\left( \begin{array}{cc} u(\alpha) & -v(\alpha) \\
\eta v(\alpha) & u(\alpha) \end{array} \right),
\eea
with $u(\alpha)$ and $v(\alpha)$ are given as
\bea
v^2(\alpha)=(e^{\alpha\omega_k}-1)^{-1}, \quad\quad u^2(\alpha)=1+v^2(\alpha).\label{phdef}
\eea
Here, the parameter $\alpha$ is defined $\alpha=(\alpha_0,\alpha_1,\cdots\alpha_{D-1})$, where $D$ is the space-time dimension. It is called the compactification parameter. It is important to note that, in this formalism, any space-time dimension can be compactified.

For the applications that follow in the next section, it is important to consider the scalar field propagator in the TFD approach. In this context, the propagator is written as
\bea
G_0^{(AB)}(x-x';\alpha)&=&i\int \frac{d^4k}{(2\pi)^4}e^{-ik(x-x')}G_0^{(AB)}(k;\alpha),
\eea
where $A, B=1, 2$ define the doubled notation and
\bea
G_0^{(AB)}(k;\alpha)={\cal B}^{-1}(\alpha)G_0^{(AB)}(k){\cal B}(\alpha),
\eea
with $G_0(k)$ being the usual massless scalar field propagator. The choice $A=B=1$ leads to the physical component, which is given by the non-tilde variables. Then
\bea
G_0^{(11)}(k;\alpha)=G_0(k)+\eta\, v^2(k;\alpha)[G^*_0(k)-G_0(k)],
\eea
where $v^2(k;\alpha)$ is the generalized Bogoliubov transformation, which is defined as \cite{GBT}
\bea
v^2(k;\alpha)=\sum_{s=1}^d\sum_{\lbrace\sigma_s\rbrace}2^{s-1}\sum_{l_{\sigma_1},...,l_{\sigma_s}=1}^\infty(-\eta)^{s+\sum_{r=1}^sl_{\sigma_r}}\,\exp\left[{-\sum_{j=1}^s\alpha_{\sigma_j} l_{\sigma_j} k^{\sigma_j}}\right],\label{BT}
\eea
with $d$ being the number of compactified dimensions and $\lbrace\sigma_s\rbrace$ denotes the set of all combinations with $s$ elements.

Using this formalism the vacuum expectation value of the energy-momentum tensor Eq. (\ref{VEV}) becomes
\bea
\left\langle T_{\mu\nu}^{(AB)}(x;\alpha)\right\rangle=\lim_{x'\rightarrow x}\Bigl\{i\Delta_{\mu\nu}G_0^{(AB)}(x-x';\alpha)-\Sigma_{\mu\nu}\delta(x-x')\delta^{(AB)}\Bigl\}.
\eea
To obtain a physical result, a renormalization procedure must be carried out, which consists of
\bea
{\cal T}_{\mu\nu}(x;\alpha)=\left\langle T_{\mu\nu}^{(AB)}(x;\alpha)\right\rangle-\left\langle T_{\mu\nu}^{(AB)}(x)\right\rangle.
\eea
This leads to
\bea
{\cal T}_{\mu\nu}(x;\alpha)=\lim_{x'\rightarrow x}\Bigl\{i\Delta_{\mu\nu}\overline{G}_0^{(AB)}(x-x';\alpha)\Bigl\},\label{EMT}
\eea
with
\bea
\overline{G}_0^{(AB)}(x-x';\alpha)=G_0^{(AB)}(x-x';\alpha)-G_0^{(AB)}(x-x').\label{Green}
\eea
Since the physical quantities are given by the non-tilde components, i.e. $A=B=1$, let us take $\overline{G}_0(x-x';\alpha)=\overline{G}_0^{(11)}(x-x';\alpha)$.

In the next section, the gravitational background, where the applications at finite temperature will be investigated, is presented.

\section{G\"{o}del-type universe}

In this section, the G\"{o}del-type metrics are introduced and their main characteristics are discussed. These metrics are solutions of Einstein field equations and their principal characteristic are the so-called Closed Time-like Curves (CTC's). An observer traveling on these curves can return to the past, at least theoretically. This leads to the violation of causality. The first G\"{o}del-type metric was proposed by Kurt G\"{o}del in 1949 \cite{Godel}. The G\"{o}del solution describes a rotating universe with non-vanishing cosmology constant and dust-like as matter source. The G\"{o}del metric is given as
\bea
ds^2=[dt+H(x)dy]^2-D^2(x)dy^2-dx^2-dz^2,\label{Godel}
\eea
where the functions  $H(x)$ and $D(x)$ are defined as
\bea
H(x)&=&e^{mx},\\
D(x)&=&\frac{e^{mx}}{\sqrt{2}}.
\eea
This metric is a solution of the Einstein field equations if the following relations are satisfied
\bea
m^2=2\omega^2=-2\Lambda=\kappa\rho
\eea
with $\omega$ being the vorticity of matter, $\Lambda$ the cosmological constant, $\kappa$ the Einstein constant and $\rho$ the energy density. The G\"{o}del-type metric is a generalization of the metric given in Eq. (\ref{Godel}). In cylindrical  coordinates the G\"{o}del-type line element is written as \cite{Reboucas}
\bea
ds^2=[dt+H(r)d\phi]^2-D^2(r)d\phi^2-dr^2-dz^2,\label{cylindrical}
\eea
where the functions $H(r)$ and $D(r)$ satisfy the relations
\bea
\frac{H'(r)}{D(r)}&=&2\omega\nonumber\\
\frac{D''(r)}{D(r)}&=&m^2.
\eea 
Here the prime denotes the derivative with respect to $r$.  The parameters $m^2$ and $\omega$ completely characterize the properties of the G\"{o}del-type metrics and take the values $\omega\neq 0$ and $-\infty\leq m^2\leq +\infty$. The signal of parameter $m^2$ defines three different classes of the G\"{o}del-type metrics:
\begin{enumerate}
\item linear class ($m^2=0$ and $\omega\neq 0$):
\bea
H(r)&=&\omega r^2,\nonumber\\
D(r)&=&r.
\eea

\item trigonometric class ($m^2<0$ and $\omega\neq 0$):
\bea
H(r)&=&\frac{2\omega}{\mu^2}[1-\cos(\mu r)],\nonumber\\
D(r)&=&\frac{1}{\mu}\sin(\mu r), 
\eea
where $m^2=-\mu^2$.

\item  hyperbolic class ($m^2>0$ and $\omega\neq 0$):
\bea
H(r)&=&\frac{4\omega}{m^2}\sinh^2\left(\frac{mr}{2}\right),\nonumber\\
D(r)&=&\frac{1}{m}\sinh(m r). 
\eea
\end{enumerate}

It is to be noted that, the G\"{o}del metric is recovered for the case $m^2=2\omega^2$, then this metric belongs to the hyperbolic class. For the applications that follow, the hyperbolic class is considered. In order to investigate the presence of CTC, the line element (\ref{cylindrical}) is written as
\bea
ds^2=dt^2+2H(r)dt d\phi - dr^2-G(r)d\phi^2 -dz^2,
\eea
where $G(r)=D^2(r)-H^2(r)$. The CTC arises if $G(r)$ is negative for a range of $r$-values ($r_1 < r < r_2$). The hyperbolic class can be divided into two regions: (i) $0<m^2<4\omega^2$ and (i) $m^2\geq 4\omega^2$. In the first case, there is a non-causal region for $r>r_c$, where the critical radius is defined as
\bea
\sinh^2\left(\frac{mr_c}{2}\right)=\left(\frac{4\omega^2}{m^2}-1\right)^{-1}.
\eea
In the second case, when $m^2=4\omega^2$ the critical radius goes to infinity, then there is no violation of causality and, therefore, no CTC occurs.

In the next section, the effects of temperature in a G\"{o}del-type universe are investigated. The TFD approach is considered.

\section{Thermal applications on a G\"{o}del-type background}

Our attention is focused on calculating the components of the energy-momentum tensor associated with a scalar field coupled to gravity at finite temperature having a G\"{o}del-type solution as a gravitational background. In order to make calculations simpler, let us consider a local set of tetrad basis $\theta^A=e^A_\mu dx^\mu$, with $e^A_\mu$ being the tetrad or vierbein. A good simplification comes from choosing the basis
\bea 
\theta^{(0)}&=&dt+H(r)d\phi, \quad \theta^{(1)}=dr,\nonumber\\
\theta^{(2)}&=&D(r)d\phi, \quad\quad \quad \theta^{(3)}=dz.
\eea
Then the G\"{o}del-type line element (\ref{cylindrical}) takes the form
\bea
ds^2=\eta_{AB}\theta^A\theta^B=(\theta^{(0)})^2-(\theta^{(1)})^2-(\theta^{(2)})^2-(\theta^{(3)})^2.
\eea
Here the capital Latin letters are labeled as tetrad indices and $\eta_{AB}$ is the Minkowski metric.

The energy-momentum tensor (\ref{EMT}) in the tetrad basis becomes
\bea
{\cal T}_{AB}(x;\alpha)=\lim_{x'\rightarrow x}\Bigl\{i\Delta_{AB}\overline{G}_0(x-x';\alpha)\Bigl\},\label{EMT2}
\eea
where
\bea
\Delta_{AB}&=&\frac{1}{2}\eta_{AB}\partial^\rho\partial'_\rho-\partial_A\partial'_B+\xi\left(R_{AB}-\frac{1}{2}\eta_{AB}R+\eta_{AB}\Box-\partial_A\partial'_B\right),
\eea
with
\bea
\eta_{AB}&=&e^\mu_A e^\nu_B g_{\mu\nu}, \quad \partial_A=e^\mu_A\partial_\mu, \quad R_{AB}=e^\mu_A e^\nu_B R_{\mu\nu}.
\eea
It is to be noted that, $e^\mu_A$ is the inverse of $e^A_\mu$ that satisfies the condition $e^A_\mu e^\mu_B=\delta^A_B$.

In the Local Lorentz frame (tetrad basis), the non-zero components of the Ricci tensor are
\bea
R_{(0)(0)}=2\omega^2\quad\quad R_{(1)(1)}=R_{(2)(2)}=2\omega^2-m^2,
\eea
and the Ricci scalar is
\bea
R=2(m^2-\omega^2).
\eea

In this framework, let us investigate different applications at finite temperature.  To achieve this goal, the TFD topological structure which is defined as $\Gamma_D^d=(\mathbb{S}^1)^d\times \mathbb{R}^{D-d}$ with $1\leq d \leq D$ is used. Here $D$ are the space-time dimensions and $d$ is the number of compactified dimensions. Thus, three situations are analyzed for different choices of the compactification parameter $\alpha$ which is defined $\alpha=(\alpha_0,\alpha_1,\cdots\alpha_{D-1})$: (i) $\alpha=(\beta,0,0,0)$ in the topology $\Gamma_4^1=\mathbb{S}^1\times\mathbb{R}^{3}$, where $\beta=1/k_B T$ with $k_B$ being the Boltzmann constant and $T$ the temperature. In this case the time-axis is compactified in $\mathbb{S}^1$, with circumference $\beta$. In this way, the effects of temperature are introduced.  (ii) $\alpha=(0,0,0,i2d)$ considering the topology $\Gamma_4^1$, with the compactification along the coordinate $z$. This leads to size effects (Casimir effect).  (iii)  $\alpha=(\beta,0,0,i2d)$  with $\Gamma_4^2=\mathbb{S}^1\times\mathbb{S}^1\times\mathbb{R}^{2}$. This implies double compactification, i.e., one being time and the other along the coordinate $z$. Such compactifications allow calculating the Casimir effect at finite temperature.

\subsection{ Thermal effects in the G\"{o}del-type universe}

In order to calculate the thermal effects, $\alpha=(\beta,0,0,0)$ is chosen. For this case, the generalized Bogoliubov transformation (\ref{BT}) and the Green function are given, respectively, as
\bea
v^2(\beta)&=&\sum_{l_0=1}^{\infty}e^{-\beta k^0l_0},\label{BT1}\\
\overline{G}_0(x-x';\beta)&=&2\sum_{l_0=1}^{\infty}G_0(x-x'-i\beta l_0n_0).\label{GF1}
\eea
Then  energy-momentum tensor Eq. (\ref{EMT2}), becomes
\bea
{\cal T}_{AB}(x;\beta)=2i\lim_{x'\rightarrow x}\Bigl\{\Delta_{AB}\sum_{l_0=1}^{\infty}G_0(x-x'-i\beta l_0n_0)\Bigl\},\label{41}
\eea
where $n_0=(1,0,0,0)$. To make the calculations of the last equation, it is necessary to use the massless scalar field propagator which is defined as
\bea
G_0(x-x')=-\frac{i}{(2\pi)^2}\frac{1}{(x-x')^2},
\eea
with
\bea
(x-x')^2&=&\eta_{AB}(x-x')^A(x-x')^B\nonumber\\
&=&\eta_{AB}e^A_\mu e^B_\nu(x-x')^\mu(x-x')^\nu \nonumber\\
&=&(t-t')^2+2H(r)(t-t')(\phi-\phi')-(r-r')^2\nonumber\\
&-&\left[D^2(r)-H^2(r)\right](\phi-\phi')^2-(z-z')^2.
\eea
Here the non-zero tetrad components, i.e. $e^{(0)}_0=e^{(1)}_1=e^{(3)}_3=1, e^{(0)}_2=H(r), e^{(2)}_2=D(r)$, have been used.

The component $A=B=0$ of Eq. (\ref{41}) leads to
\bea
E(T)=\frac{\pi^2}{30}(1+\xi)T^4+\frac{\xi(m^2-3\omega^2)}{12}T ^2,\label{ST}
\eea
where $E(T)\equiv{\cal T}_{(0)(0)}(T)$. This is the Stefan-Boltzmann law for the scalar field coupled to gravity in a G\"{o}del-type universe. It is to be noted that, the first term is the usual and dominant contribution at high temperatures, while the second term represents the modification due to the gravitational background and becomes relevant at low temperatures. In addition, there are two cases that must be considered. (i) Choosing $m^2=2\omega^2$ leads to the G\"{o}del universe. In this case, there is a critical temperature for which the second term of Eq. (\ref{ST}), i.e. $-\frac{\xi\omega^2}{12}T ^2$, becomes dominant. This implies that a negative energy density arises. (ii) For the choice $m^2=4\omega^2$ which implies a causal G\"{o}del-type solution, the second term is always positive, i.e. $\frac{\xi\omega^2}{12}T ^2$. Therefore, in a causal universe, the energy density is positive for any temperature value as expected.

\subsection{Size effect at zero temperature}

Here $\alpha=(0,0,0,i2d)$ is chosen. Then the Casimir effect at zero temperature is calculated. To develop such a calculation, it is considered
\bea
v^2(d)=\sum_{l_3=1}^{\infty}e^{-i2d k^3l_3}\label{BT2}
\eea
as the Bogoliubov transformation and 
\bea
\overline{G}_0(x-x';d)=2\sum_{l_3=1}^{\infty}G_0(x-x'-2d l_3n_3)\label{GF2}
\eea
with $n_3=(0,0,0,1)$ is the Green function. Thus the energy-momentum tensor (\ref{EMT2}) is given as
\bea
{\cal T}_{AB}(x;d)=2i\lim_{x'\rightarrow x}\Bigl\{\Delta_{AB}\sum_{l_3=1}^{\infty}G_0(x-x'-2d l_3n_3)\Bigl\}.
\eea

From this expression, the Casimir energy and pressure are obtained choosing, respectively, $A=B=0$ and $A=B=3$. Then
\bea
{\cal T}_{(0)(0)}(d)&=&-\frac{\pi^2}{1440d^4}(1+\xi)+\xi\frac{(m^2-3\omega^2)}{48d^2}, \\
{\cal T}_{(3)(3)}(d)&=&-\frac{\pi^2}{480d^4}(1+\xi)-\xi\frac{(m^2-\omega^2)}{48d^2}.\label{CEZ}
\eea
It is important to note that the Casimir energy and pressure are negative, which implies an attractive force. Therefore, the Casimir force is attractive for both causal and non-causal G\"{o}del-type solutions. Furthermore, the question of causality does not change the nature of this phenomenon.

\subsection{Effects of temperature and size in a G\"{o}del-type universe}

In order to investigate effects due to the temperature and spatial compactification, the $\alpha$ parameter is chosen as  $\alpha=(\beta,0,0,i2d)$. For this case, the generalized Bogoliubov transformation is composed of three parts:  the first one corresponds to the Stefan-Boltzmann law given in Eq. (\ref{BT1}), the second is associated with the Casimir effect at zero temperature Eq. (\ref{BT2}) and the third part is formed as a combination of the first two parts. Then
\bea
v^2(\beta,d)=\sum_{l_0=1}^\infty e^{-\beta k^0l_0}+\sum_{l_3=1}^\infty e^{-i2dk^3l_3}+2\sum_{l_0,l_3=1}^\infty e^{-\beta k^0l_0-i2dk^3l_3}.\label{BT3}
\eea
To calculate the Casimir effect at finite temperature, the Green function related to the third part of the Bogoliubov transformation is needed. It is given as
\bea
\overline{G}_0(x-x';\beta,b)&=&4\sum_{l_0,l_3=1}^\infty G_0\left(x-x'-i\beta l_0n_0-2dl_3n_3\right).\label{GF3}
\eea
Using Eq. (\ref{GF3}) in Eq. (\ref{EMT2}) and taking $A=B=0$ we get
\bea
{\cal T}_{(0)(0)}(\beta,d)&=&-\frac{1}{\pi^2}\sum_{l_0,l_3=1}^\infty\frac{1}{[(2dl_3)^2+(\beta l_0)^2]^3}\Bigl\{(1+\xi)[2(2dl_3)^2-6(\beta l_0)^2]\nonumber\\
&-&\xi\left[(m^2-3\omega^2)\left((2dl_3)^4+(\beta l_0)^4+2(2dl_3)^2(\beta l_0)^2\right)\right]\Bigl\}.
\eea
This is the Casimir energy at finite temperature in a G\"{o}del-type universe. Choosing $A=B=3$ leads to
\bea
{\cal T}_{(3)(3)}(\beta,d)&=&-\frac{1}{\pi^2}\sum_{l_0,l_3=1}^\infty\frac{1}{[(2dl_3)^2+(\beta l_0)^2]^3}\Bigl\{(1+\xi)[6(2dl_3)^2-2(\beta l_0)^2]\nonumber\\
&+&\xi\left[(m^2-\omega^2)\left((2dl_3)^4+(\beta l_0)^4-2(2dl_3)^2(\beta l_0)^2\right)\right]\Bigl\},\label{CET}
\eea
which is the Casimir pressure at finite temperature. Here the behavior of pressure at non-zero temperature must be analyzed. As discussed in the previous subsection, the Casimir effect is attractive at zero temperature for both causal and non-causal G\"{o}del-type universes. However, at finite temperature this situation is different, that is, there is a critical temperature where the Casimir pressure is zero. From this value up to high temperatures, the Casimir effect becomes repulsive. Figure 1 shows this behavior. Furthermore, this result occurs for both causal and non-causal G\"{o}del-type universes. In addition, the critical temperature increases with the value of the $\omega^2$ parameter. In Figures 2 and 3, the behavior of pressure versus temperature is shown  for different values of $\omega^2$ for the cases $m^2=2\omega^2$ and $m^2=4\omega^2$, respectively. Therefore, it is shown how the critical temperature varies with the parameter $\omega^2$. It is important to emphasize that, the parameters $d$ and $\xi$, are assumed to be fixed, since $d=10^{-6}m$ is the separation between the plates used in some experiments \cite{Lamoreaux, Mohideen} and $\xi=1/6$ is a usual value for the coupling constant \cite{Birrel, Mota}.
\begin{figure}[h]
\includegraphics[scale=0.9]{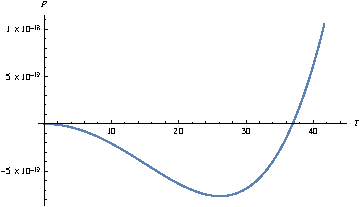}
\caption{Casimir pressure ($P={\cal T}_{(3)(3)}(T)$) as a function of the temperature $T$. It is used $d\sim \mu m$ ($\mu=10^{-6}$) \cite{Lamoreaux, Mohideen}, $\xi=1/6$ \cite{Birrel, Mota} and $m^2=4\omega^2$ with $\omega^2=10^3$.}
\end{figure}

\begin{figure}[h]
\includegraphics[scale=0.7]{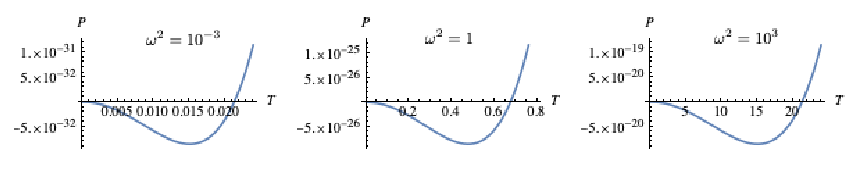}
\caption{Non-causal solution ($m^2=2\omega^2$) - Casimir pressure versus temperature for different values of $\omega^2$.}
\end{figure}

\begin{figure}[h]
\includegraphics[scale=0.7]{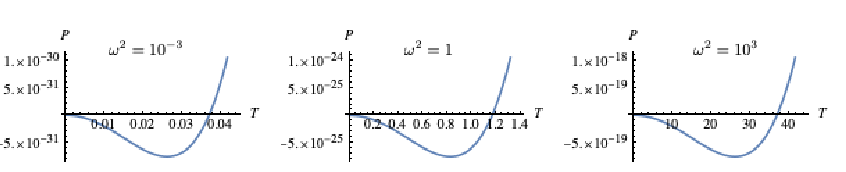}
\caption{Causal solution ($m^2=4\omega^2$) - Casimir pressure versus temperature for different values of $\omega^2$.}
\end{figure}

\section{Conclusions}

One of the most fascinating quantum phenomena is the Casimir effect. It occurs for any quantum field as a consequence of the vacuum fluctuations of that field under specific boundary conditions or topological effects. To investigate this phenomenon, a massless scalar field coupled to gravity is considered. For such development, the TFD formalism is used. It is a real-time quantum field theory at a finite temperature that is built from two fundamental elements: the doubling of the Hilbert space and the Bogoliubov transformation. Here its topological structure is explored allowing the study of different effects such as temperature effect and size effect in the same way. Such phenomena are studied having the G\"{o}del-type solutions as the gravitational background. G\"{o}del-type metrics can display CTCs that lead to a causality violation. Then our main objective is to analyze whether this causality breakdown affects the Stefan-Boltzmann law and the Casimir effect for a massless scalar field coupled to gravity. The first result is obtained in Eq. (\ref{ST}), i.e. the Stefan-Boltzmann law, which shows that the G\"{o}del universe must be discarded to avoid negative energies. However, it is possible that another region in this space is a causal region and the energy density is always positively defined. Another result is Eq. (\ref{CEZ}) which exhibits that the Casimir effect at zero temperature in G\"{o}del-type universes is attractive for both causal and non-causal regions. The main result of this paper is given in Eq. (\ref{CET}). It is the thermal Casimir effect in the G\"{o}del-type universes. It is important to note that, as shown in Figures 1, 2 and 3, there is a critical temperature where the Casimir effect goes to zero. Then from this point up to high temperatures the Casimir effect becomes repulsive. This behavior arises in both causal and non-causal regions. Therefore, for a certain temperature, there is a phase transition from an attractive Casimir effect to a repulsive Casimir effect. This transition from the attractive to the repulsive Casimir effect has been also obtained in other physical systems, as an example see the reference \cite{our2}.

\section*{Acknowledgments}

This work by A. F. S. is partially supported by National Council for Scientific and Technological Develo\-pment - CNPq project No. 313400/2020-2. 


\global\long\def\link#1#2{\href{http://eudml.org/#1}{#2}}
 \global\long\def\doi#1#2{\href{http://dx.doi.org/#1}{#2}}
 \global\long\def\arXiv#1#2{\href{http://arxiv.org/abs/#1}{arXiv:#1 [#2]}}
 \global\long\def\arXivOld#1{\href{http://arxiv.org/abs/#1}{arXiv:#1}}


\end{document}